\documentclass[sort&compress
    ,final            
  ]
  {aipproc}

\layoutstyle{8x11single}
   \usepackage{showkeys}
    \usepackage{amssymb}
    \usepackage{bm}
    \usepackage{amsmath}

\newcommand\beq{\begin{equation}}
\newcommand\eeq{\end{equation}}
\newcommand\beqa{\begin{eqnarray}}
\newcommand\eeqa{\end{eqnarray}}

\begin{document}

\title{Strong breakdown of equipartition in uniform granular mixtures
}


\author{Andr\'es Santos}{
  address={Departamento de F\'{\i}sica, Universidad de Extremadura, Badajoz,
  Spain}
}
\author{James W. Dufty}{
  address={Department
of Physics, University of Florida, Gainesville, FL 32611} }



\begin{abstract}
A binary mixture made of heavy  and light  inelastic hard spheres in
the free cooling state is considered. First, the regions in the
parameter space where the partial granular temperature of the heavy
species  is larger (or smaller) than that of the light species are
analyzed. Next, the asymptotic behavior of the mean square velocity
ratio in the disparate-mass limit is investigated, assuming
different scaling laws  for the parameters of the mixture. It is
seen that two general classes of states are possible: a ``normal''
state and an ``ordered'' state, the latter representing a strong
breakdown of energy equipartition.

\end{abstract}

\maketitle


\section{Introduction}
As is well known, granular fluids are intrinsically out of
equilibrium. This clearly manifests itself in the case of granular
mixtures, where the equipartition of energy is broken down (even in
homogeneous and isotropic states), as shown by theory
\cite{GD99,BT02a}, simulations \cite{simu}, and experiments
\cite{exp}. In this work we consider a binary mixture made of heavy
($h$) and light ($\ell$) inelastic hard spheres in the free cooling
state. By using kinetic theory tools, one can derive a tenth-degree
algebraic equation \cite{GD99} whose solution gives the mean square
velocity ratio $\phi=\langle v_h^2\rangle/\langle
v_\ell^2\rangle=T_h m_\ell/T_\ell m_h$ (which plays the role of an
``order'' parameter) as a function of the control parameters of the
problem (the densities, the three coefficients of normal
restitution, and the mass and size ratios).  We first investigate
the regions in the parameter space where the partial granular
temperature of the heavy species is larger ($T_h> T_\ell$) or
smaller ($T_h< T_\ell$) than that of the light species. We next
analyze the asymptotic behavior of $\phi$ in the disparate-mass
limit ($m_\ell/m_h\to0$), assuming different scaling laws for the
coefficients of restitution, size ratio, and concentrations. We
observe that it is possible to distinguish two general classes of
states: a \textit{normal} state (where $\phi\to 0$) and an
\textit{ordered} state (where $\phi\neq 0$), the latter representing
a strong breakdown of energy equipartition in which the temperature
ratio $T_h/T_\ell$ diverges. These two classes can be further
resolved. Thus, the normal state can be \textit{partitioned}
($T_h/T_\ell=\text{finite}$), \textit{mono-energetic $h$}
($T_h/T_\ell\to\infty$), or \textit{mono-energetic $\ell$}
($T_h/T_\ell\to 0$). Moreover, if the concentration of  heavy
particles vanishes, the ordered state can become
\textit{super-ordered} ($\phi\to\infty$). The phase diagrams
corresponding to this rich phenomenology are presented. In
particular, in the limit of vanishing concentration of  heavy
particles, the partitioned phase is separated from the ordered and
from the super-ordered phases by two critical lines. This extends a
previous analysis made in the case of an impurity particle immersed
in an inelastic gas \cite{SD01}.

\section{Basic equations and conditions for $T_h>T_\ell$}

Let us consider the homogeneous cooling state of a binary mixture of
two granular species ($h$ and $\ell$) characterized by mole
fractions $x_h=n_h/n$ and $x_\ell=1-x_h$, masses $m_h$ and
$m_\ell<m_h$, diameters $\sigma_{h}$, $\sigma_{\ell}$, and
$\sigma_{h\ell}=(\sigma_{h}+\sigma_{\ell})/2$, coefficients of
normal restitution $\alpha_{hh}$, $\alpha_{\ell \ell}$, and
$\alpha_{h\ell}$, and pair correlation contact values $\chi_{hh}$,
$\chi_{\ell \ell}$, and $\chi_{h\ell}$. This gives 10 independent
parameters. However, in appropriate reduced units the number of
control parameters reduces considerably.  The set of two Boltzmann
equations are
\beq
\partial_t f_h(\mathbf{v})=J_{hh}[\mathbf{v}|f_h,f_h]+J_{h\ell}[\mathbf{v}|f_h,f_\ell], \quad
\partial_t f_\ell(\mathbf{v})=J_{\ell h}[\mathbf{v}|f_\ell,f_h]+J_{\ell \ell}[\mathbf{v}|f_\ell,f_\ell].
\label{0.2}
\eeq
Multiplying  by $v^2$ and integrating over velocity one gets
\beq
\partial_t \langle v_h^2\rangle =-(\zeta_{hh}+\zeta_{h\ell}) \langle v_h^2\rangle,
\quad \partial_t \langle v_\ell^2\rangle =-(\zeta_{\ell
\ell}+\zeta_{\ell h}) \langle v_\ell^2\rangle,\quad
\zeta_{ij}=\frac{\int d\mathbf{v}\, v^2
J_{ij}[\mathbf{v}|f_i,f_j]}{\int d\mathbf{v}\, v^2 f_i(\mathbf{v})}.
\label{0.11}
\eeq
The diagonal terms  $\zeta_{hh}$ and $\zeta_{\ell \ell}$ are
positive definite and represent the inelastic cooling rates of
species $h$ and $\ell$, respectively.  On the other hand, the cross
term $\zeta_{h\ell}$ represents the ``thermalization'' rate of
species $h$ due to collisions with particles of species $\ell$ and
can be either positive or negative, depending on the state of the
mixture. Analogously, the term $\zeta_{\ell h}$  represents the
thermalization rate of species $\ell$ due to collisions with
particles of species $h$.

The four rates $\zeta_{ij}$ are complicated nonlinear functionals of
the unknown distribution functions $f_h$ and $f_\ell$. However,
reasonably good estimates can be obtained by using the Maxwellian
approximation
\beq
f_i(\mathbf{v})=n_i \left({m_i}/{2\pi
T_i}\right)^{3/2}\exp\left(-{m_iv^2}/{2T_i}\right),\quad
T_i={m_i}\langle v_i^2\rangle/3.
\label{0.4}
\eeq
This approximation yields  $\zeta_{ij}=\omega \xi_{ij}$, where
$\omega={4\pi}n \chi_{h\ell}\sigma_{h\ell}^2\langle v_\ell\rangle
({1+\alpha_{h\ell}}){m_\ell}/3({m_h+m_\ell})$ is an effective
collision frequency and the dimensionless cooling and thermalization
rates are \cite{GD99}
\beq
\xi_{hh}(\phi)=x_h\sqrt{\phi}\beta_h,\quad \xi_{h\ell}(\phi)=x_\ell
\frac{\sqrt{1+\phi}}{1+\phi_0}\left(1-\mu+\phi_0-\frac{\mu}{\phi}\right),
\label{3}
\eeq
\beq
\xi_{\ell \ell}(\phi)=x_\ell\beta_\ell,\quad \xi_{\ell h}(\phi)=x_h
\frac{\sqrt{1+\phi}}{1+\phi_0}\left(1+\frac{\phi_0-\phi}{\mu}+\phi\right)(1-\mu).
\label{4}
\eeq
Here, $\phi \equiv {\langle v_h^2\rangle}/{\langle v_\ell^2\rangle}$
is the mean square velocity ratio, $\mu\equiv
{m_\ell}/({m_h+m_\ell})$ is the mass ratio,
$\phi_0\equiv({1-\alpha_{h\ell}})/({1+\alpha_{h\ell}})$ is a measure
of the cross coefficient of restitution, and the coefficients
\beq
\beta_h\equiv\frac{1+\phi_0}{4\sqrt{2}}\frac{1-\alpha_{hh}^2}{\mu}\frac{\chi_{hh}}{\chi_{h\ell}}
\left(\frac{\sigma_{h}}{\sigma_{h\ell}}\right)^2,\quad
\beta_\ell\equiv\frac{1+\phi_0}{4\sqrt{2}}\frac{1-\alpha_{\ell
\ell}^2}{\mu}\frac{\chi_{\ell \ell}}{\chi_{h\ell}}
\left(\frac{\sigma_{\ell}}{\sigma_{h\ell}}\right)^2
\label{2}
\eeq
essentially  measure the self coefficients of restitution and are
inversely proportional to $\mu$.
 The evolution equations \eqref{0.11} yield $
\omega^{-1}\partial_t \phi=(\xi_{\ell \ell}+\xi_{\ell
h}-\xi_{hh}-\xi_{h\ell})\phi$, so that the homogeneous cooling state
of the mixture is characterized by the stationarity condition
\beq
\xi_{hh}(\phi)+\xi_{h\ell}(\phi)-\xi_{\ell \ell}(\phi)-\xi_{\ell
h}(\phi)=0.
\label{cond}
\eeq
This equation can be transformed into a tenth-degree algebraic
equation for $\phi$. Its physical solution determines $\phi$ as a
function of the five control parameters of the problem, namely
$\{x_h,\alpha_{hh},\alpha_{\ell \ell},\alpha_{h\ell},m_\ell/m_h\}$
 or, equivalently, $\{x_h,\beta_h,\beta_\ell,\phi_0,\mu\}$.

Equipartition of energy implies that $\phi=\mu/(1-\mu)$. Of course,
this happens when all the collisions are elastic (i.e.,
$\beta_h=\beta_\ell=\phi_0=0$). When collisions are inelastic, the
most common situation is  $\phi>\mu/(1-\mu)$, i.e., the heavy
particles have a larger (partial) temperature than the light
particles, $T_h>T_\ell$ \cite{simu,exp}. However, this is not always
necessarily the case. To clarify this point, it is convenient to
define the parameters $K_i\equiv \mu\beta_i(1+\phi_0)/\phi_0$,
$i=h,\ell$, which essentially measure the degree of inelasticity of
the $i$-$i$ collisions, relative to the inelasticity of the
$h$-$\ell$ collisions. A careful analysis of Eq.\ \eqref{cond} shows
that, for a given value $\mu<\frac{1}{2}$, one has $T_h>T_\ell$ when
one of the following three alternative cases occurs: (i)
$K_h>K_h^*(\mu)$, $K_\ell>K_\ell^*(\mu)$, and
$x_h<x_h^*(K_h,K_\ell,\mu)$; (ii) $K_h<K_h^*(\mu)$,
$K_\ell<K_\ell^*(\mu)$, and $x_h>x_h^*(K_h,K_\ell,\mu)$; (iii)
$K_h<K_h^*(\mu)$ and $K_\ell>K_\ell^*(\mu)$. Here,
\beq
K_h^*(\mu)\equiv\frac{1-\mu}{\sqrt{\mu}},\quad K_\ell^*(\mu)\equiv
\frac{\mu}{\sqrt{1-\mu}},\quad x_h^*(K_h,K_\ell,\mu)\equiv
\frac{K_\ell \sqrt{1-\mu}-\mu}{K_h\sqrt{\mu}+K_\ell \sqrt{1-\mu}-1}.
\label{x.10}
\eeq

\begin{figure}[ht]
\includegraphics[width=\columnwidth]{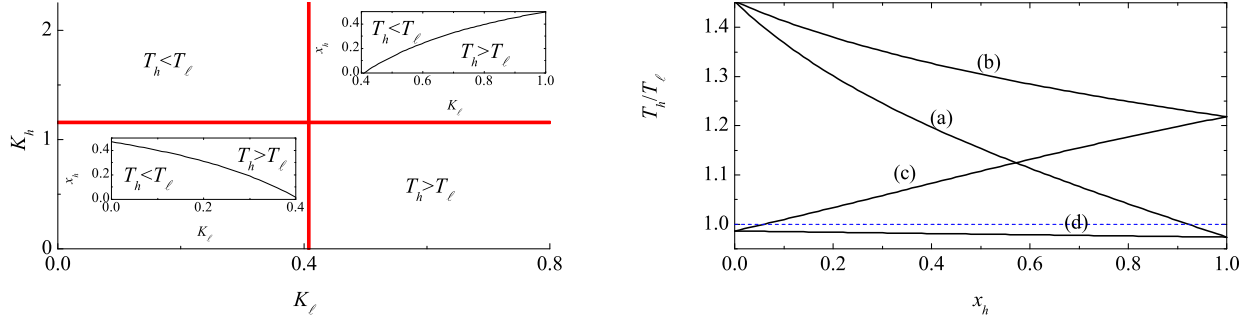}
\caption{Left panel: Plane $K_h$ vs $K_\ell$ split into four
quadrants by the lines $K_h=K_h^*(\mu)$ and $K_\ell=K_\ell^*(\mu)$
for $\mu=\frac{1}{3}$; in the top-right quadrant, $T_h>T_\ell$ if
$x_h<x_h^*(K_h,K_\ell,\mu)$, as represented by the inset, where
$x_h^*$ is plotted as a function of $K_\ell$ for $\mu=\frac{1}{3}$
and $K_h=2$; in the bottom-right quadrant, $T_h>T_\ell$ for any
concentration; in the bottom-left quadrant, $T_h>T_\ell$ if
$x_h>x_h^*(K_h,K_\ell,\mu)$, as represented by the inset, where
$x_h^*$ is plotted as a function of $K_\ell$ for $\mu=\frac{1}{3}$
and $K_h=0.5$; finally, in the top-left quadrant, $T_h<T_\ell$ for
any concentration. Right panel: Plot of the temperature ratio
$T_h/T_\ell$ versus $x_h$ for $\alpha_{h\ell}=0.8$, $m_h=2m_\ell$,
$\sigma_{h}=\sigma_{\ell}$, and (a) $(\alpha_{hh},\alpha_{\ell
\ell})=(0.6,0.6)$, (b) $(\alpha_{hh},\alpha_{\ell \ell})=(0.9,0.6)$,
(c) $(\alpha_{hh},\alpha_{\ell \ell})=(0.9,0.9)$, and  (d)
$(\alpha_{hh},\alpha_{\ell \ell})=(0.6,0.9)$.
\label{example}}
\end{figure}
The left panel of Fig.\ \ref{example} sketches the regions in the
$K_h$--$K_\ell$ plane   where $T_h>T_\ell$ or $T_h<T_\ell$ for a
given mass ratio. The right panel shows the temperature ratio versus
concentration for a few representative cases.

\section{The disparate-mass limit. Phase diagrams}
 We are now interested in the limit $m_\ell/m_h\to
0\Rightarrow \mu\to 0$. The question we want to address is, how does
the solution $\phi$ of Eq.\ \eqref{cond} behave in the limit $\mu\to
0$? There are two distinct possibilities: either $\lim_{\mu\to
0}\phi=0$ or $\lim_{\mu\to 0}\phi\neq 0$. {}Thus, $\phi$ can be
viewed as an ``order'' parameter. The  case $\lim_{\mu\to 0}\phi=0$
means that the heavy particles move much more slowly than the light
particles; we will say that the mixture is then in a \textit{normal}
state. The alternative case $\lim_{\mu\to 0}\phi\neq 0$ implies that
the speed of the heavy particles is typically comparable (or even
larger) than that of the light species; we will call
\textit{ordered} to such a state.

 In order to make a finer classification, let us introduce a ``critical exponent'' $\eta$ such that
$\phi\sim \mu^\eta$ as $\mu\to 0$, so $\eta>0$ corresponds to normal
states and $\eta\leq 0$ correspond to ordered states. If $\eta=1$,
the equipartition of energy is weakly broken in the sense that
$T_h/T_\ell=\text{finite}$, as in the elastic case. This corresponds
to a subclass of the normal state that we will call
\textit{partitioned}. An example of this situation occurs when the
self-collisions are elastic and the cross collisions are
quasi-elastic, with $1-\alpha_{h\ell}\sim \mu$. On the other hand,
if $\eta>1$, then $\phi$ decays even more rapidly than in the
elastic limit. Again, this is a normal state ($\phi\to 0$) but is
not partitioned ($T_h/T_\ell\to 0$) since all the energy is carried
by the light particles. This case will be referred to as a
\textit{mono-energetic $\ell$} state, an example of which taking
place when only the $h$-$h$ collisions are inelastic. Still in the
normal class, the case $0<\eta<1$  represents a breakdown of
equipartition intermediate between that of the partitioned state and
that of the ordered state: while the heavy particles move much more
slowly than the light ones, they carry all the energy of the
mixture. We will refer to this situation as a \textit{mono-energetic
$h$} state. A simple example corresponds to mixtures of disparate
sizes when only the $\ell$-$\ell$ collisions are inelastic. Thus,
three subclasses (mono-energetic $\ell$, partitioned, and
mono-energetic $h$) can be distinguished within the class of normal
states. As for ordered states, the typical case corresponds to
$\eta=0$, i.e., $\phi=\text{finite}$ in the disparate-mass limit.
This happens, for instance, when only  the cross collisions are
inelastic, regardless of the values of the mole fractions. In
particular, the state is ordered both in the ``Brownian'' limit
$x_h\to 0$ and in the ``Lorentz'' limit $x_\ell\to 0$. In the former
case, $\phi\neq 0$ implies that the \textit{dynamics} is not
Brownian since the solute Brownian particles move with a speed
comparable to that of the fluid particles. In the latter case, the
system does not behave as a Lorentz gas since the heavy
``scatterers'' are never at rest. An even stronger breakdown of
energy equipartition occurs if $\eta<0$, implying not only $\phi\neq
0$ (ordered state), but $\phi\to \infty$. We will say that in this
case the system reaches a \textit{super-ordered} state. A necessary
condition for the existence of this situation is  the Brownian limit
$x_h\to 0$. A super-ordered state appears, for instance, if only the
$\ell$-$\ell$ collisions are inelastic and $x_h\sim\mu$, so that the
total masses of both species are comparable.

\begin{table}[ht]
\begin{tabular}{llcccc}
\hline
\tablehead{1}{l}{b}{Class}&\tablehead{1}{l}{b}{Subclass}&\tablehead{1}{c}{b}{$\eta$}
&\tablehead{1}{c}{b}{$\langle v_h^2\rangle/\langle v_\ell^2\rangle$}&\tablehead{1}{c}{b}{$T_h/T_\ell$}&\tablehead{1}{c}{b}{Example}\\
\hline
Normal&Mono-energetic $\ell$&$\eta>1$&0&0&$\alpha_{hh}<1$\\
&Partitioned&$\eta=1$&0&finite&$1-\alpha_{h\ell}\sim m_\ell/m_h$\\
&Mono-energetic $h$&$0<\eta<1$&0&$\infty$&$\alpha_{\ell \ell}<1, \sigma_{i}\sim m_i^{1/3}$\\
\hline
Ordered&Ordered&$\eta=0$&finite&$\infty$&$\alpha_{h\ell}<1$\\
&Super-ordered&$\eta<0$&$\infty$&$\infty$&$\alpha_{\ell \ell}<1,
x_h\sim m_\ell/m_h$\\
\hline
\end{tabular}
\caption{Possible classes of states in a binary granular mixture in
the disparate-mass limit $m_\ell/m_h\approx \mu\to 0$. The mean
square ratio $\phi=\langle v_h^2\rangle/\langle v_\ell^2\rangle$
scales as $\phi\sim \mu^\eta$.}
\label{table0}
\end{table}
Table \ref{table0} describes the five possible subclasses (or
phases) and the corresponding representative examples. In general,
depending on the regime of values for the control parameters
$\alpha_{ij}$ and $x_h$, the asymptotic homogeneous cooling state of
a disparate-mass binary mixture belongs in one of those subclasses.
In order to investigate the regions in the parameter space
corresponding to each phase, let us assume scaling laws of the form
\beq
\beta_h\sim \mu^{-1+a_h},\quad \beta_\ell\sim \mu^{-1+a_\ell},\quad
\phi_0\sim \mu^b, \quad x_h\sim \mu^{c_h},
\label{5.1}
\eeq
where the exponents $a_h$, $a_\ell$, $b$, and $c_h$ are
non-negative. The exponents $a_h$ and $b$ measure, respectively, the
inelasticity of the $h$-$h$ and $h$-$\ell$ collisions (i.e.,
$1-\alpha_{hh}\sim \mu^{a_h}$, $1-\alpha_{h\ell}\sim \mu^{b}$),
while $a_\ell$ measures the inelasticity of the $\ell$-$\ell$
collisions, as well as the possible dependence of the size ratio on
the mass ratio: $(1-\alpha_{\ell \ell})(\sigma_\ell/\sigma_h)^2\sim
\mu^{a_\ell}$. If any of the three types of collision is elastic,
the corresponding exponent takes an infinite value. Conversely, if
the collision is inelastic then the exponent vanishes. A finite
value of $a_h$ or $b$ implies that the associated collision is
quasi-elastic. In the case $a_\ell=\text{finite}$, either the
$\ell$-$\ell$ collisions are quasi-elastic or the size ratio scales
with the mass ratio, or both. For finite concentrations of the $h$
species one has $c_h=0$, while $c_h>0$ in the Brownian limit
($x_h\to 0$).  {}From Eqs.\ \eqref{3} and \eqref{4}, the scaling
behaviors of $\xi_{ij}$ become
\beq
\xi_{hh}\sim \mu^{-1+a_h+c_h+\eta/2},\quad \xi_{h\ell}\sim
\text{max}\{\mu^0,\mu^{\eta/2}\}\text{max}\{\mu^0,|\mu^{1-\eta}|\},
\label{5.2}
\eeq
\beq
\xi_{\ell \ell}\sim \mu^{-1+a_\ell},\quad \xi_{\ell h}\sim
\mu^{c_h}\text{max}\{\mu^0,\mu^{\eta/2}\}\text{max}\{\mu^0,\mu^{-1+b},|\mu^{-1+\eta}|\}.
\label{5.3}
\eeq
In the above equations, $|\cdots|$ indicates that the corresponding
term has a negative prefactor.

\begin{figure}[ht]
\includegraphics[width=\columnwidth]{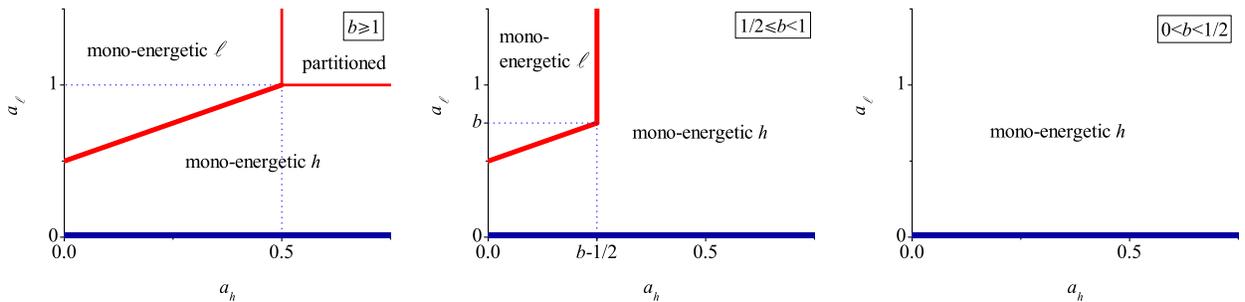}
\caption{Regions in the plane $a_\ell$ vs $a_h$ corresponding to
each phase. The thick lines separating the phases mono-energetic
$\ell$ and mono-energetic $h$ correspond to partitioned states. The
thick line at $a_\ell=0$ represents ordered states. The whole plane
is occupied by ordered states if $b=0$ (not shown).
\label{phase_diagram2}}
\end{figure}

Let us first assume that the mole fraction $x_h$ is kept finite in
the limit $\mu\to 0$, so that $c_h=0$. In that case, condition
\eqref{cond} is fulfilled in the limit $\mu\to 0$ when any of the
five conditions described in Table \ref{table1} is satisfied.
 The associated constraints and expressions for the exponent
$\eta$ are also included  in Table \ref{table1}.  It is easy to
check that in region B one always has $\eta>1$, while in region D
one always has $0<\eta<1$. In region E, $\eta=1$  if $b\geq 1$,
while $0<\eta<1$  if $0<b<1$. As for region A, $0<\eta<1$
 if $a_\ell<a_h+\frac{1}{2}$, while $\eta>1$
 otherwise, this latter case being only possible
when $b\geq\frac{1}{2}$. Likewise, in region C one has $0<\eta<1$
 if $a_h>b-\frac{1}{2}$ and $\eta>1$
 otherwise, this latter case being only possible
when $b\geq\frac{1}{2}$. The phase diagrams are shown in Fig.\
\ref{phase_diagram2}.
\begin{table}[ht]
\begin{tabular}{lllc}
\hline
\tablehead{1}{c}{b}{Case}&\tablehead{1}{c}{b}{Conditions}&\tablehead{1}{c}{b}{Constraints}&\tablehead{1}{c}{b}{$\eta$}\\
\hline
A&$\xi_{hh}\sim\xi_{\ell \ell}\gg |\xi_{h\ell}|,|\xi_{\ell h}|$&$2a_h< a_\ell<\text{min}\{b,\frac{2}{3}(a_h+1)\}$&$2(a_\ell-a_h)$\\
B&$\xi_{hh}\sim|\xi_{h\ell}|\gg \xi_{\ell \ell},|\xi_{\ell
h}|$&$a_\ell> \frac{2}{3}(a_h+1)$,
$a_h<\text{min}\{\frac{1}{2},\frac{3}{2}b-1\}$&$\frac{2}{3}(2-a_h)$\\
C&$\xi_{hh}\sim|\xi_{\ell h}|\gg \xi_{\ell
\ell},|\xi_{h\ell}|$&$b<1$, $a_\ell> b$,
$\frac{3}{2}b-1< a_h< \frac{1}{2}b$&$2(b-a_h)$\\
D&$\xi_{\ell \ell}\sim|\xi_{\ell h}|\gg \xi_{hh},|\xi_{h\ell}|$&$ a_\ell< \text{min}\{2a_h,b,1\}$&$a_\ell$\\
E&$|\xi_{\ell h}|\gg \xi_{\ell \ell},
\xi_{hh},|\xi_{h\ell}|$&$a_h>\frac{1}{2}\text{min}\{b,1\}$,
$a_\ell>\text{min}\{b,1\}$&$\text{min}\{b,1\}$\\
\hline
\end{tabular}
\caption{Possible cases in the disparate-mass limit at finite
concentration ($c_h=0$).}
\label{table1}
\end{table}

\begin{figure}[ht]
\includegraphics[width=0.5\columnwidth]{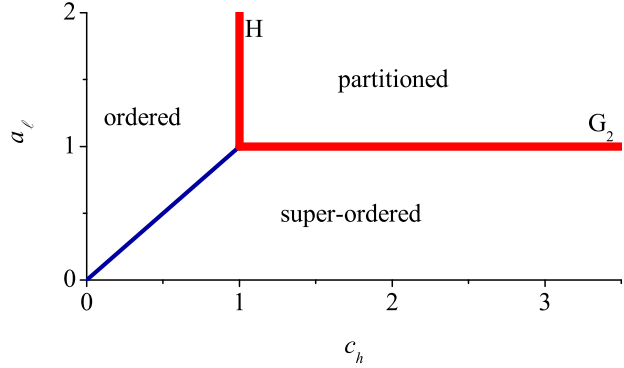}
\caption{Regions in the plane $a_\ell$ vs $c_h$ corresponding to the
phases present in the limit of vanishing concentration of heavy
particles when $1-\alpha_{h\ell}$ is kept finite ($b=0$). The thick
line separating the regions super-ordered and partitioned (line
G$_2$) and the thick line separating the regions ordered and
partitioned (line H) are critical lines on which the state can be
partitioned, mono-energetic $h$, or ordered, depending on the value
of the parameters $\beta_\ell$ (line G$_2$) or $\gamma$ (line H).
\label{phase_diagram4}}
\end{figure}

Now we assume that the concentration of $h$ particles behaves as
$x_h\sim \mu^{c_h}$ in the limit $\mu\to 0$. The total mass of the
$h$ species is larger than that of the $\ell$ species (i.e., $x_h
m_h\gg x_\ell m_\ell$) if $c_h<1$, while the opposite happens if
$c_h>1$. This limit adds a new parameter $c_h>0$ to the parameter
space, so that the phase diagrams in Fig.\ \ref{phase_diagram2} are
modified. Let us restrict ourselves to $b=0$, i.e., the cross
collisions are kept inelastic in the limit $\mu\to 0$. As indicated
in the caption of Fig.\ \ref{phase_diagram2}, if $b=0$ the state of
the mixture is always ordered ($\eta=0$) in the case of finite
concentrations ($c_h=0$). However, this uniform situation changes
dramatically if $c_h>0$. With $b=0$, it is easy to realize from
Eqs.\ \eqref{5.2} and \eqref{5.3} that $|\xi_{\ell h}|$ is never
negligible versus $\xi_{hh}$. That means that the value of the
parameter $a_h$ (i.e., the degree of inelasticity of the $h$-$h$
collisions) will not play any role, so the parameter space reduces
to $a_\ell$ and $c_h$. The  possible cases are described in Table
\ref{table2}.  Cases G$_1$ and G$_2$ correspond to
$\text{max}\{\mu^0,\mu^{\eta/2}\}=\mu^{\eta/2}$ and
$\text{max}\{\mu^0,\mu^{\eta/2}\}=\mu^0$, respectively. While G$_1$
defines a whole region in  the parameter space with a well-defined
(negative) expression for the exponent $\eta$, G$_2$ defines a line
along which the value of $\eta$ is not unique. A similar situation
occurs with case H. The phase diagram is  shown in Fig.\
\ref{phase_diagram4}. We observe that $\eta$ changes discontinuously
from $\eta=0$ to $\eta=1$ when going from the ordered  region to the
partitioned region through line H. Even more discontinuous is the
transition from the super-ordered region ($\eta<0$) to the
partitioned region ($\eta=1$) through line G$_2$. These are
\textit{critical} lines along which the value of $\eta$ is not
uniquely determined by the values of $a_\ell$ and $c_h$. Let us
analyze these critical lines with some detail.
\begin{table}[ht]
\begin{tabular}{lllc}
\hline
\tablehead{1}{c}{b}{Case}&\tablehead{1}{c}{b}{Conditions}&\tablehead{1}{c}{b}{Constraints}&\tablehead{1}{c}{b}{$\eta$}\\
\hline
F&$|\xi_{h\ell}|\gg \xi_{\ell \ell},|\xi_{\ell h}|$&$a_\ell> 1$, $c_h> 1$&$1$\\
G$_1$&$|\xi_{h\ell}|\sim\xi_{\ell \ell}\gg |\xi_{\ell
h}|$&$\frac{1}{2}(3-c_h)< a_\ell< 1$,
$c_h> 1$&$2(a_\ell-1)$\\
G$_2$&$|\xi_{h\ell}|\sim\xi_{\ell \ell}\gg |\xi_{\ell
h}|$&$a_\ell=1$, $c_h> 1$&$0\leq\eta\leq
1$\\
 H&$|\xi_{h\ell}|\sim |\xi_{\ell h}|\gg \xi_{\ell \ell}$&$c_h=1$, $a_\ell>1$&$0\leq \eta\leq 1$\\
I&$\xi_{\ell \ell}\sim|\xi_{\ell h}|\gg |\xi_{h\ell}|$&$a_\ell<\text{min}\{c_h,\frac{1}{2}(3-c_h)\}$&$\frac{2}{3}(a_\ell-c_h)$\\
J&$|\xi_{\ell h}|\gg \xi_{\ell \ell}, |\xi_{h\ell}|$&$a_\ell> c_h$,
$c_h< 1$&$0$\\
\hline
\end{tabular}
\caption{Possible cases in the disparate-mass limit with $x_h\to 0$
and $\alpha_{h\ell}<1$ ($b=0$).}
\label{table2}
\end{table}

Along the critical line G$_2$  the leading terms are $\xi_{h\ell}$
and $\xi_{\ell \ell}$, so that condition \eqref{cond} becomes
\beq
{\sqrt{1+\phi}}\left[1-\frac{\mu}{\phi(1+\phi_0)}\right]={\beta_\ell}.
\label{x.11}
\eeq
The solution to this equation in the limit $\mu\to 0$ is
\beq
\phi=
\begin{cases}
\frac{1}{1-\beta_\ell}\frac{\mu}{1+\phi_0}&\text{if }\beta_\ell<1,\\
\sqrt{\frac{2\mu}{1+\phi_0}}&\text{if }\beta_\ell=1,\\
\beta_\ell^2-1&\text{if }\beta_\ell>1.
\end{cases}
\label{x.12}
\eeq
Therefore, $\beta_\ell$ plays the role of a \textit{control}
parameter and $\beta_\ell=1$ is its critical value. If
$\beta_\ell<1$ the state is partitioned, if $\beta_\ell=1$ it is
mono-energetic $h$, and if $\beta_\ell>1$ it is ordered. This
situation has been analyzed in detail in Ref.\ \cite{SD01}. Since
$c_h> 1$, the total mass of the $h$ species is negligible versus
that of the $\ell$ species. Moreover, the $\ell$ particles are not
affected by the presence of the $h$ particles.

Along the critical line H, $x_h\sim \mu$ (i.e., $c_h=1$), what means
that both species have comparable total masses. In addition, the
$\ell$-$\ell$ collisions are not too inelastic (or the size ratio
$\sigma_{\ell}/\sigma_{h}$ is sufficiently small), in the sense that
$(1-\alpha_{\ell
\ell}^2)(\sigma_{\ell}/\sigma_{h})^2\sim\mu^{a_\ell}$ with
$a_\ell>1$. In that situation, the leading terms are $\xi_{h\ell}$
and $\xi_{\ell h}$. Therefore, condition \eqref{cond} yields
\beq
1-\frac{\mu}{\phi(1+\phi_0)}=\frac{x_h}{\mu}\frac{\phi_0-\phi}{1+\phi_0},
\label{x.13}
\eeq
whose solution is
\beq
\phi=
\begin{cases}
\frac{1}{1-\gamma}\frac{\mu}{1+\phi_0}&\text{if }\gamma<1,\\
\sqrt{\frac{\phi_0\mu}{1+\phi_0}}&\text{if }\gamma=1,\\
\frac{\phi_0}{\gamma}(\gamma-1)&\text{if }\gamma>1.
\end{cases}
\label{x.15}
\eeq
Here, $\gamma\equiv
({x_h}/{\mu}){\phi_0}/{(1+\phi_0})=({x_h}/{\mu})({1-\alpha_{h\ell}})/{2}$,
which represents the control parameter in this case.

\section{Conclusions}
In this paper we have shown that in the homogeneous cooling state of
a granular binary mixture the mean square velocity ratio
$\phi=\langle v_h^2\rangle/\langle v_\ell^2\rangle$ and the
temperature ratio $T_h/T_\ell$ can take widely different values
depending on the parameters of the system (coefficients of
restitution, concentrations, and mass and size ratios). Typically,
$T_h>T_\ell$ if (i) the self-collisions ($h$-$h$ and $\ell$-$\ell$)
are sufficiently more inelastic than the cross collisions
($h$-$\ell$) and the concentration of the $h$ particles is
sufficiently low; (ii) the $h$-$h$ and $\ell$-$\ell$ collisions are
sufficiently less inelastic than the $h$-$\ell$ collisions and $x_h$
is large enough; (iii) the $h$-$h$ collisions are not too inelastic
but the $\ell$-$\ell$ collisions are sufficiently inelastic.

In the disparate-mass limit $\mu\to 0$, the breakdown of
equipartition can become extreme with divergent, finite, or
vanishing limits of the mean square velocity and/or temperature
ratios, depending on the scaling behavior of the parameters. This
gives rise to the five classes of states (or phases) described in
Table \ref{table0}, ranging from the mono-energetic $\ell$ state
($T_h/T_\ell\to 0$) to the super-ordered state ($\phi\to\infty$).
The associated phase diagrams are presented in Figs.\
\ref{phase_diagram2} and \ref{phase_diagram4}. If the cross
collisions are inelastic ($\alpha_{h\ell}<1$), the state is always
ordered ($\phi=\text{finite}$). As a consequence, in this case there
is neither Brownian dynamics (when $x_h\to 0$) nor Lorentz dynamics
(when $x_\ell\to 0$). A partitioned state
($T_h/T_\ell=\text{finite}$) is only possible if the three types of
collisions are sufficiently quasi-elastic, while a super-ordered
state ($\phi\to\infty$) is only possible in the Brownian limit
($x_h\to 0$). In that limit, there is no mono-energetic $\ell$ state
and there exist critical lines in the phase diagram where the state
can be partitioned, ordered, or mono-energetic $h$.

The analysis performed in this work has been restricted to the
homogeneous cooling state, but it can be easily extended to mixtures
heated with a white-noise forcing \cite{BT02a}. In that case,
condition \eqref{cond} is replaced by
$\phi\left[\xi_{hh}(\phi)+\xi_{h\ell}(\phi)\right]-\xi_{\ell
\ell}(\phi)-\xi_{\ell h}(\phi)=0$. As a consequence, the phase
monoenergetic $\ell$ is suppressed.


\begin{theacknowledgments}
A.S. acknowledges partial support from the Ministerio de Educaci\'on
y Ciencia
 (Spain) through Grant No.\ FIS2004-01399.
\end{theacknowledgments}



\bibliographystyle{aipproc}   

\end{document}